\documentclass[
]{ceurart}

\usepackage{listings}
\begin{document}

\copyrightyear{2025}
\copyrightclause{Copyright for this paper by its authors.
  Use permitted under Creative Commons License Attribution 4.0
  International (CC BY 4.0).}

\conference{DaQuaMRec Workshop at RecSys ’25, September 22–26, 2025, Prague, Czech Republic}

\title{Data Quality Challenges in Multimodal Tourism Recommender Systems}

\author[1,2]{Zehui Wang}[%
orcid=0000-0001-5401-8773,
email=Zehui.Wang@rwu.de,
]
\cormark[1]
\address[1]{University of Applied Sciences Ravensburg-Weingarten, Doggenriedstrasse, Weingarten, Weingarten,  88250, Germany}
\address[2]{University of Klagenfurt, Universitätsstraße 65–67, Klagenfurt am W{\"o}rthersee, 9020, Austria}

\author[1]{Wolfram H{\"o}pken}[%
orcid=0000-0002-4175-1295,
email=Wolfram.Hoepken@rwu.de,
]

\author[2]{Dietmar Jannach}[%
orcid=0000-0002-4698-8507,
email=Dietmar.Jannach@aau.at,
]

\cortext[1]{Corresponding author.}

\begin{abstract}
  Modern recommender systems increasingly consider diverse data modalities to enhance model performance. However, the quality of the underlying multimodal data (e.g. data from social media platforms) has received limited attention so far. Focusing on tourism recommendation, we examine real-world challenges encountered when using datasets from Yelp and Instagram. Issues include noise from fake reviews, hallucinated content from LLMs, modality contradictions, and demographic bias. We outline future directions such as quality-aware fusion, structured and controlled LLM-based content generation to reduce hallucinations, and user group differentiation to improve the robustness and reliability of multimodal recommendation systems.
\end{abstract}


\maketitle

\section{Motivation}
As a key application domain of recommender systems, tourism recommendation involves tasks such as providing personalized suggestions for transportation, accommodation, and points-of-interest (POIs)~\cite{Ricci2020Recommender,Borras2014Intelligent,Sarkar2023Tourism}. Recently, researchers have increasingly explored the use of diverse information types to improve model performance~\cite{Wang2025ASurvey,Zhang2024ASurvey}. Beyond check-in records that incorporate timestamps and geographic coordinates, additional sources such as POI profiles (e.g., textual attributes), social relationships (e.g., friendship networks), user feedback (e.g., reviews), and visual content (e.g., photos) have been integrated through representation learning, attention-based neural architectures, or late fusion schemes~\cite{Wang2025ASurvey}.

Despite the growing reliance on such heterogeneous data sources, the quality of multimodal information remains largely underexplored. This oversight can introduce noise, bias, and contradictions into the recommendation process. In the following, we draw on findings from real-world datasets to illustrate the challenges that arise when working with imperfect, incomplete, or biased multimodal data in tourism recommender systems.

\section{Challenges in Multimodal Data Quality}
Recent investigations in tourism-related user behavior analysis and next-POI recommendation have revealed several fundamental issues regarding the quality of multimodal data. Based on two real-world datasets from~\cite{Wang2025DoTravel,wang2025beyond}, namely the Multimodal Yelp Dataset\footnote{\url{https://huggingface.co/datasets/wzehui/Yelp-Multimodal-Recommendation}} and an Instagram dataset collected from the Lake Constance region in Germany\footnote{This dataset cannot be publicly shared due to platform restrictions and privacy considerations.}, we summarize the key challenges as follows.

\paragraph*{Noise in User-Generated and Model-Generated Content.}

Real-world datasets often contain noisy information that can significantly degrade the effectiveness of recommendation models. For example, the multimodal Yelp dataset is constructed from user-generated reviews and check-in records on the Yelp platform. 
However, some of these reviews originate from fake or automated accounts that produce fabricated interactions, leading to distorted representations of businesses. Prior work \cite{wang2025beyond} has tackled this issue by analyzing behavioral patterns to identify non-human users. Suspicious accounts are first flagged based on unusually high posting frequency compared to normal user activity. These flagged accounts are then further examined for implausible movement speeds between consecutive check-ins and eventually removed. In the Instagram dataset, burst postings (multiple posts within a few minutes at the same location) are also observed. Retaining such data does not contribute to the analysis of meaningful sequential behavior patterns, but instead dilutes the signal of genuine user trajectories and biases the statistical distribution of activities.

In addition, with the increasing use of LLMs to process textual and visual modalities, a new form of semantic noise has emerged. Hallucinated or irrelevant content generated by LLMs can significantly degrade the quality of downstream embeddings and harm recommendation performance. One approach to address this issue is to use structured summarizations instead of open-ended generation. By enforcing predefined fields in the summaries, the process becomes more controllable and less susceptible to hallucination. In particular, for visual content, summaries are generated selectively, only for images that contain clear and interpretable visual cues \cite{wang2025beyond}. Ambiguous images are filtered out to avoid introducing additional noise.

\paragraph*{Demographic Bias in User Behavior and Content Distribution.}

Demographic bias occurs when there is a mismatch between the distribution of users in training data and the characteristics of the target population \cite{NeophytouRevisiting2022}. A typical case in tourism is the behavioral divergence between tourists and locals. For example, Instagram posts collected from Lake Constance region includes contributions from both travelers and nearby residents. If such data are used indiscriminately, the recommendation model may overfit to local preferences that are irrelevant for tourist-oriented next-POI suggestions.

Demographic bias can also arise from population imbalances in content creation. For instance, Instagram content may primarily come from younger users, whereas recommendations targeting regions like Lake Constance are often aimed at an older audience. 
This mismatch can reduce the relevance and effectiveness of recommendations. Moreover, the majority of media originates from the DACH region (Germany, Switzerland, Austria), reflecting cultural and behavioral patterns specific to this geographic area, which limits the generalizability of the dataset to broader international contexts. One possible approach to mitigate this issue is to 
further analyze the user profiles to 
improve audience segmentation and better align the system output with the preferences of the target user group.

\paragraph*{Semantic Contradictions Across Modalities.}

Contradictions refer to semantic inconsistencies between content from different modalities. In the current datasets, such inconsistencies mainly stem from user-generated content (e.g., misaligned captions, incorrect location tags, irrelevant hashtags) or from business-provided information that is outdated or inaccurate. A common example arises in the multimodal Yelp dataset, where an LLM-generated summary based on the majority of user reviews might describe a restaurant as ``clean and quiet'', while user-uploaded photos reveal a cluttered or noisy environment. Similar inconsistencies can also be found in Instagram posts, where images, captions, and location tags do not always align. A photo may be taken in one city but tagged to another, or hashtags may be added that are unrelated to the actual content. Such contradictions reduce the coherence and reliability of the fused multimodal representation, posing challenges for both modality alignment and inference stability in downstream models.

\section{Possible Solutions \& Future Directions}
Given such challenges, we propose several directions for future research:
(1) Future models should adopt data-aware multimodal learning, where the quality of each modality is explicitly modeled during training. This way, low-quality modalities can be dynamically downweighted in the modality fusion process \cite{ZhangProvable2023,WeiRobust2024};
(2) Current fusion strategies assume consistent and complete input across modalities, which rarely holds in real-world scenarios. Future research should develop robust fusion mechanisms to tolerate missing, noisy, or redundant inputs \cite{LinMissModal2023,TsaiMultimodal2020,CaiTrusted2024}. This includes support for preserving modality-specific uniqueness, reducing cross-modal redundancy, and enhancing synergistic signals to better adapt to imperfect data;
(3) Future studies should also focus on modeling user intent divergence across different groups, such as tourists versus locals. Identifying and representing such behavioral differences during preprocessing or representation learning can mitigate distributional mismatches and improve recommendation relevance \cite{SanchezTravelers2022,Derdouri2021};
(4) In addition to these directions, we observe that videos (e.g., ones posted by social media users or 360° immersive marketing videos)~\cite{Casillo2025Integrating,Argyriou2020} have not been leveraged much in the literature for improved tourism recommendations. We see the incorporation of these additional visual signals as a promising area for future work;
(5) Given the increasing role of LLMs in multimodal recommendation, we advocate for more research on structured and controlled LLM-based generation to reduce hallucinations, thereby enhancing the reliability of generated content for downstream use \cite{king-etal-2022-dont}. 

In summary, integrating data quality considerations into both upstream preprocessing and downstream modeling is essential for building trustworthy, personalized, and data-aligned multimodal recommender systems.


%

\bibliography{PositionPaper}


\end{document}